\documentclass[]{spiejour}  %>>> use for US letter paper 
\usepackage{amsmath}  %>>> for AMS math formatting, 
\usepackage[]{graphicx}
\usepackage{epstopdf,mathrsfs,amsmath,amsfonts}
\DeclareMathOperator{\erf}{erf}
\DeclareMathOperator{\sgn}{sgn}
\newcommand{\logmine}[1]{\textrm{$\ \textrm{log}_{10}\textrm{(#1)}$}}
\newcommand{\etal}{\textit{et al.}}
\newtheorem{postulate}[theorem]{Postulate}
\newtheorem{corollary}[theorem]{Corollary}

\title{Kramers-Kr\"{o}nig analysis of biological skin}

\author{Luisiana X. Cundin\supscr{a} and William P. Roach\supscr{b}}

\affiliation{\supscr{a}Conceptual Mindworks Inc., 9830 Colonnade Blvd, San Antonio, TX  78230\\
\linkable{luisiana.cundin@gmail.com} \\
\supscr{b}USAF AFMC AFRL/RDLA, 3550 Aberdeen Ave SE, Kirtland AFB, NM 87117\\
\linkable{william.roach@kirtland.af.mil}
}

\begin{document} 
  \maketitle 

%%%%%%%%%%%%%%%%%%%%%%%%%%%%%%%%%%%%%%%%%%%%%%%%%%%%%%%%%%%%% 
\begin{abstract}
A treatise on the optical property of biological tissue is presented. Water is postulated to be a topological basis and serves to discriminate published skin data. Electromagnetic theory governing dielectric behavior is concisely detailed pertaining to certain optical constants and Kramers-Kr\"{o}nig relation. The Kramers-Kr\"{o}nig relation defining dispersion index is emulated through the discrete Hilbert transform. An accrued absorption set is populated with empirical absorption data for biological skin, pure liquid water and interpolated values. Kramers-Kr\"{o}nig analysis of biological skin yields a comprehensive description of the complex index of refraction from DC to x-ray frequencies.   
\end{abstract}

%>>>> Include a list of up to six keywords after the abstract

\keywords{Kramers-Kr\"{o}nig, Hilbert, absorption, dispersion index}

%%%%%%%%%%%%%%%%%%%%%%%%%%%%%%%%%%%%%%%%%%%%%%%%%%%%%%%%%%%%%

\section{Introduction}
A survey of published optical constants for biological skin underscores attention has been placed on the Radio Frequency (RF), near-infrared (NIR) and visible (VIS) frequency bands. Attention has shifted recently to the terahertz band (THz), due to newly available radiation sources. Within each of these bands, reported values can differ dramatically from author to author and are commonly attributed to inherent error in experimental technique, numerical model used to calculate optical constants or random error. Experimentalists are limited to the narrow frequency band of their source and often cannot comprehend how their data fits into the broader picture. Theory can provide a comprehensive perspective of dielectric behavior, bring empirical data into focus and illuminate misconceptions. 

Biological skin is a multi-layered heterogeneous dielectric material. Three typical divisions of skin are epidermis, dermis and subdermis layers, where epidermis is frequently referenced as the \emph{stratum corneum}. The partial molar configuration varies for each distinct layer of skin and also across racial groups, changes with age and is affected by disease, to name a few factors. Enumeration of so many variables discourages any unified treatment of biological skin. A global approach becomes accessible if water is postulated as a topological basis for all skin types. Consequently, it becomes possible to treat biological skin as an electromagnetic continuum.  

Electromagnetic theory describes the interaction of electric and magnetic fields with ponderable media through constitutive relations. Each constitutive relation must be known for the bandwidth of radiation considered to accurately model and predict electrodynamic behavior of biological skin. Theory cannot produce results independent of experimental data. Electromagnetic theory permits calculating dispersion index from known absorption behavior through the well-known Kramers-Kr\"{o}nig relation \cite{Landau}. Kramers-Kr\"{o}nig analysis of aggregate tissue absorption data yields a comprehensive dispersion curve spanning the entire frequency spectrum.    

\section{Theory and Methods}
Maxwell's equations, combined with all appropriate constitutive relations, fully describe electromagnetic behavior in ponderable media \cite{Landau,Jackson}. There are three relevant constitutive relations for electromagnetic phenomena in dielectric media; the permittivity, admittance and permeability. Each relation is represented as a complex function of frequency and position, where dependence on position can be neglected in isotropic media. The complex permeability ($\hat{\mu}$) determines magnetization of a material in response to externally applied magnetic fields. The complex admittance ($\hat{\sigma}$) relates applied fields to induced current densities and is the sum of conductivity ($\sigma'$) and susceptivity ($\sigma''$). Conductivity measures the ease direct current flows through a material and susceptivity measures the ease dynamic current changes direction. The permittivity ($\hat{\epsilon}$) measures the influence of an electric field to charge migration and dipole reorientation. The real and imaginary parts of the complex permittivity describe stored energy in the system ($\epsilon'$) and a dissipative term ($\epsilon''$) measuring the loss of energy to the system.
\begin{equation}\label{constitutivecond}
\begin{array}{c}
\hat{\epsilon}=\epsilon'+i\epsilon'' \\
\hat{\mu}=\mu'+i\mu''\\
\hat{\sigma}=\sigma'+i\sigma''
\end{array} 
\end{equation}

The complex index of refraction ($\hat{\textrm{N}}$) combines all three constitutive relations into one complex function \cite{Jackson,Segelstein}. Because the complex index of refraction is a dimensionless function, each constitutive relation is relative to vacuum, where the vacuum permittivity is $\epsilon_0\approx 8.854\times 10^{-12}\ \textrm{F/m}$ and vacuum permeability is $\mu_0=4\pi\times 10^{-7}\ \textrm{H/m}$. For non-magnetic material, the relative permeability is considered equal to unity. The dispersion index (\textit{n}) measures frequency specific variation in phase velocity and accounts for reflection, refraction, chromatic dispersion and elastic scattering of incident electromagnetic radiation. Both Rayleigh and Mie scattering are examples of elastic processes causing redirection of electromagnetic energy. The absorption (\textit{k}) term measures loss of energy to the system, including nonconservative losses through inelastic scattering processes. 
\begin{equation}\label{complexrefraction}
 \hat{\textrm{N}}(\omega)=n+ik=\sqrt{\hat{\mu}_r \hat{\epsilon}_r+i\frac{4\pi 
\hat{\mu}_r\hat{\sigma}_r}{\omega}}
\end{equation}

Electromagnetic theory states absorption is null as frequency passes through the origin and infinite point \cite{Landau,Jackson}. Equation (\ref{limitofnk}) shows the one-sided limit for dispersion index and absorption in terms of elementary constitutive relations, where $a=\epsilon'_r-4\pi\sigma''_r/\omega,b=\epsilon''_r+4\pi\sigma'_r/\omega$. Invoking Minkowski's inequality to approximate the inner square root term found in each definition reduces each one-sided limit, see Theorem \ref{Minkowski}.  The dispersion index is directly proportional to real permittivity, referenced as the static permittivity or static dielectric constant. Absorption is shown primarily dependent upon the imaginary permittivity. This property of absorption enables an accurate calculation for low frequencies knowing only conductivity, where the following relation exists: $\epsilon''=\sigma'/\omega$. 
\begin{equation}\label{limitofnk}
 \lim_{\omega\rightarrow 0^+}\left\{
\begin{array}{c}
n(\omega)=\sqrt{(a+\sqrt{a^2+b^2})/2} \\
k(\omega) =\sqrt{(-a+\sqrt{a^2+b^2})/2}
\end{array} \right\}\rightarrow
\left\{
\begin{array}{c}
  \sqrt{a}>0\\
  \sqrt{b/2}\Rightarrow 0
\end{array} \right\}
\end{equation}

\begin{theorem}[Minkowski's inequality]\label{Minkowski}
The following inequality exists for two real numbers a and b $\sim$
 \begin{equation}
 \left(a^2+b^2\right)^{1/2}\leq a+b,\ \{a,b\in\mathbb{R}\}
\end{equation}
This relation becomes an equality if either a or b equal zero.
\end{theorem}

In the limit of $\omega\rightarrow\infty$, the dispersion index approaches unity, while absorption rapidly approaches zero \cite{Landau,Jackson}. The physical reason for this behavior is that polarization has not enough time to respond to rapidly changing fields $\vec{E}$ and the processes responsible for differentiation of the induction field $\vec{D}$ do not occur. The frequency where this behavior becomes valid is referred to as the plasma frequency ($\omega_p$) and for light elements occurs somewhere in the far-ultraviolet spectra \cite{Landau}. For frequencies higher than the plasma frequency, the dispersion index is solely dependent upon the concentration of conducting electrons, thus, the uniqueness of two similar materials become negligible. 
\begin{corollary}[Non-uniqueness]\label{highlimit}
 The uniqueness of biological skin will fade in the limit of infinite frequency, hence, the absorption characteristics will approach that of pure liquid water. See also Postulate \ref{waterpostulate}.
\end{corollary}

In general, the processes represented by term (\textit{a}) tend to mutually cancel in the definition for absorption, but not in the case for dispersion index. Except for DC and infinite frequencies, the dispersion index is explicitly dependent upon all physical processes occurring in a material, including absorptive processes. Experimentalists employ various theoretical models to calculate a material's optical property based on measured reflectance and transmittance data. The total attenuation ($\mu_t=\mu_s+\mu_a$) of incident radiation through a material is identified as transmittance and is the combination of both the scattering coefficient ($\mu_s$) and Lambert's absorption coefficient ($\mu_a$). From measured reflectance data and theoretical model employed, the magnitude of each coefficient is determined. This is a difficult proposition, especially for turbid media like biological tissue. For lower frequencies, effects such as electrode polarization and Maxwell-Wagner effect obscure measured results \cite{Pethig:dielectric,Schwan,Barlea}. For optical frequencies, experimental methods and theoretical models are often found inadequate \cite{Pethig:passive,Yang,Prahl,Pickering}. 

Absorption can be isolated experimentally, where the total attenuation can be simplified to Lambert's absorption coefficient ($\mu_a$) alone. Since both absorption and dispersion index are spatially dependent, careful consideration of sample thickness and measured reduction in radiation intensity passing through the sample can be modeled with Beer-Lambert's law \cite{Ingle}. Lambert's absorption coefficient has a simple relationship to absorption ($\mu_a=4\pi k/\lambda$), where $\lambda$ is the wavelength of radiation. The dispersion index shares no equally simple relationship in terms of fundamental constitutive relations. 

It is for the above reasons that some reliable method is needed to generate the index of refraction for biological skin. Fortunately, electromagnetic theory provides an analytic transform defining the dispersion index \cite{Landau,Jackson,Segelstein}. The Kramers-Kr\"{o}nig relation defining the dispersion index (\textit{n}) in terms of absorption (\textit{k}) is shown in equation (\ref{KKeqn}). This relation holds for any bound complex analytic function in the upper half plane. Analyticity in the upper half of the complex plane implies \emph{causality} \cite{Bracewell,Landau,Jackson}. Formula (\ref{KKeqn}) is of particular importance, for it makes the calculation of the dispersion index, even approximately, from the absorption function. On the other hand, the uniqueness of the absorption function in terms of dispersion index is not guaranteed, for the condition $n(\omega)-1>0$ is not satisfied in general. Therefore, implementation of Kramers-Kr\"{o}nig transform must be applied to absorption. 
\begin{equation}\label{KKeqn}
n(\omega)-1=\frac{1}{\pi}
\int_{-\infty}^{\infty}{\frac{k(\omega')}{\omega'-\omega}d\omega'}\equiv -\frac{1}{\pi\omega}\ast k(\omega)
\end{equation}

The limits of integration demand knowing the behavior of the absorption function for the entire real line. Causal systems are represented mathematically by Hermitian functions, which are comprised of an even real part and an odd imaginary part. Since the absorption function is bounded and odd, then by theory, the absorptive behavior of any material is null as the frequency passes through the origin and the infinite point. The absorption behavior for any material in the negative half-plane can be constructed by taking the negative of the positive half-plane and reversing the order along the abscissa.

The convolution form of the Kramers-Kr\"{o}nig relation is shown in equation (\ref{KKeqn}), where convolution is signified by the asterisk symbol ($\ast$) \cite{Bracewell}. In convolution form, the formula is more commonly referred to as the Hilbert transform. The Hilbert transform is most conveniently evaluated with the aid of Fourier transforms. Applying the Fourier transform to a convolution results in the multiplication of the Fourier transform of each function separately. Converting the convolution integral in equation (\ref{KKeqn}) to the transform domain requires taking the forward Fourier transform ($\mathscr{F}$) of the discrete absorption values for biological skin and then multiplying by the sign function ($\sgn$), which is then multiplied by the imaginary number ($i=\sqrt{-1}$) \cite{Bracewell}. The sign function is defined in equation (\ref{signfunction}) and is the Fourier transform of the Hilbert transform kernel ($-1/\pi\omega$), where the transform takes $\omega \mapsto s$. To recover the dispersion index, unity is added after the inverse Fourier transform ($\mathscr{F}^{-1}$) is applied to the product.
\begin{equation}\label{hilbert}
n(\omega)=1+\mathscr{F}^{-1}\hspace{-1pt}\{i\sgn(s)\mathscr{F}\hspace{-1pt}\{k(\omega)\}\}
\end{equation}

The sign function is defined to be $\sim$
\begin{equation}
    \sgn(s) =  \left\{
\begin{array}{rl}
 -1, & s < 0\\
  0, & s = 0\\
  1, & s > 0
\end{array} \right.
\label{signfunction}
\end{equation}

The Hilbert transform ($\mathscr{H}$) of a Gaussian function is shown in equation (\ref{HilbertGaussian}) and is a revealing identity. Since the absorption peak can generally be represented by a Gaussian function, the identity equals the real ($\Re$) part of the product between a Gaussian and error function (erf) with imaginary argument. The typical behavior for anomalous dispersion is controlled by the relaxation time constant ($\tau$), which also controls the width of the absorption peak. Both the prominence and steepness of the dispersion curve increases with increasing relaxation time. 
\begin{equation}\label{HilbertGaussian}
 \mathscr{H}\hspace{-2pt}\left\{\exp(-\tau x^2)\right\}= \Re\hspace{-1.5pt}\left\{i\exp(-\tau x^2)\erf\hspace{-2pt} \left( i\sqrt{\tau}x  \right) \right\}
\end{equation} 

Confusion may exist concerning the role relaxation time has with regard to frequency response and the shape of the anomalous dispersion curve. Consider the Similarity Theorem from Fourier theory, a dilation of the relaxation time corresponds to a contraction in the transform domain, see Theorem \ref{similarity} \cite{Bracewell}. The relaxation time is dilated for bound water; consequently, a contraction of the thermal absorption peak is expected in the frequency domain for bound water, which leads to Corollary \ref{bound}. Conversely, the Hilbert transform has a one to one correspondence for the relaxation time, suggesting a more prominent anomalous dispersion centered about the thermal resonance for bound water, see Corollary \ref{onetoone}.
\begin{theorem}[Similarity]\label{similarity}
 The dilation of a constant $\tau$ corresponds with a contraction in the Fourier transform domain, where $t\mapsto \omega$.
\begin{equation}
 \mathscr{F}\hspace{-2pt}\left\{f(\tau\,  t)\right\}=\textrm{F}\hspace{-1pt}\left(\omega/\tau\right)/\tau
\end{equation}

\end{theorem}
\begin{corollary}[bound water]\label{bound}
 The relaxation time is dilated for bound water; consequently, a contraction of the thermal resonance peak for bound water and biological tissue is expected.
\end{corollary}
\begin{corollary}[anomalous dispersion]\label{onetoone}
 The relaxation time has one to one correspondence for the Hilbert transform; consequently, the anomalous dispersion is expected to be more prominent around the thermal resonance frequency for biological tissue.
\end{corollary}

To ensure perfect reconstruction, Shannon's \emph{sampling theorem} demands a band-limited function is sampled at least twice the rate of the highest frequency component in the signal \cite{Shannon}. Otherwise, the signal will be \emph{under-sampled}, which means fine structure represented by high frequency components will be poorly reconstructed. It is equivalent to applying a low pass filter to a signal. Due to limitations in computer resources, \emph{under-sampling} is often unavoidable; nevertheless, \emph{under-sampled} data can produce acceptable results \cite{Bracewell}. 

Since the discrete Fourier transform (DFT) will be utilized to evaluate the Hilbert transform, proper consideration must be given to the interval sampled and the placement of windows \cite{Bracewell}. The DFT is cyclical in nature and requires sampling the entire interval bounding a band-limited signal or a complete cycle or multiple of cycles for a periodic function; otherwise, boundary errors will be introduced into the calculation. The DFT also requires elements sampled at regular subintervals. Since experimental data is irregular, interpolation techniques must be employed to properly implement the Kramers-Kr\"{o}nig transform. 
%Interpolation is accurate and stable if the interval between known data points is small. Mathematica's built-in interpolation (\textbf{Interpolation[]}) algorithm will be utilized when interpolating within small intervals \cite{Mathematica}. 

Large intervals of missing empirical data present a more difficult problem. Short of producing empirical data, there are three common methods used to attack large gaps in experimental data: 1) generating data points from a theoretical model, 2) supplanting absorption data from a similar material or 3) interpolation. No theoretical models exist for biological tissue enabling accurate absorption predictions for arbitrary frequency. Supplanting data from another material is prohibited for pure substances. In the case of heterogeneous materials, supplanting data is permissible for substances sharing similar chemical composition; therefore, would naturally share similar absorption characteristics. Lastly, interpolation is accurate and stable if the interval between known data points is small, but this method is problematic for larger intervals.
 
Despite variation in concentration, water is present in all types of skin and represents a considerable percentage of biological tissue's composition. We postulate that biological tissue should share common absorption characteristics with pure liquid water, especially in the high frequency limit, see Corollary \ref{highlimit}. Also, we recognize that tissue should diverge from the ideal absorption behavior of pure liquid water for very important theoretical reasons \cite{Landau,Bracewell,Schwan,Vogel}. An additional absorption peak should appear in the absorption spectra due to electrophoresis of ions in solution. Also, the absorption peak centered around the thermal resonance should contract, see Theorem \ref{similarity} and Corollary \ref{bound}. The presence of melanin, hemoglobin and proteins should increase absorption in the visible and ultraviolet spectra \cite{Vogel}. 

\begin{postulate}[Topological basis]\label{waterpostulate}
 Pure liquid water is an element of the basis set $\mathcal{B}$ defining biological skin.
\end{postulate}

A basis $\mathcal{B}$ for topological space $\mathcal{X}$ generates a unique topology and generates alternate topologies when combined with other open sets \cite{Lee}. The addition or variation of constituents in tissue is equivalent to a collection of sets comprised of the basis $\mathcal{B}$ set intersecting a series of open sets $s\in\mathcal{S}$, where $\mathcal{S}\subset\mathcal{X}$. New topologies are generated for each unique collection and maps all possible tissues, for every animal.

The accuracy of interpolation decreases proportionately to the interval width \cite{Burden}. Increasing the interpolation order (\textit{p}) does not necessarily remedy the situation, as order \textit{p} increases so does the implied order of derivatives increase. Inherent inaccuracies in data can lead to wildly oscillating results when attempting high order interpolations. Richardson's extrapolation algorithm takes a set of approximations and through a clever algebraic relation reduces the overall order of error to produce one accurate approximation \cite{Burden}.
For a set of known data points $\{x_0,\ldots,x_i,\ldots,x_{n}\}$, Neville's iterated interpolation algorithm produces an approximation $\tilde{x}$ with implicit interpolation order $p_n$ \cite{Burden}. The interpolation order is directly related to the number ($n+1$) of known data points used to form the approximation. For $N$ approximations generated with sequentially increasing interpolation orders \{$p_1,\ldots,p_N\}$, we propose applying Richardson's extrapolation algorithm to produce a more accurate approximation. The exact source of error need not be specified to implement Richardson's algorithm, it is only required that the error be stepwise dependent, which we assume related to the interpolation order $p$. This implies an order $N$ for Richardson's method. All algorithms were written and executed in Wolfram's Mathematica software installed on an Intel\textregistered\ Core\texttrademark 2 CPU T700 @ 2 GHz computer running Microsoft\textregistered\ Windows XP\texttrademark\ with 2 GB of RAM.

\section{Experiment}
\begin{figure}[ht]
\centering
 \includegraphics[width=5.0 in]{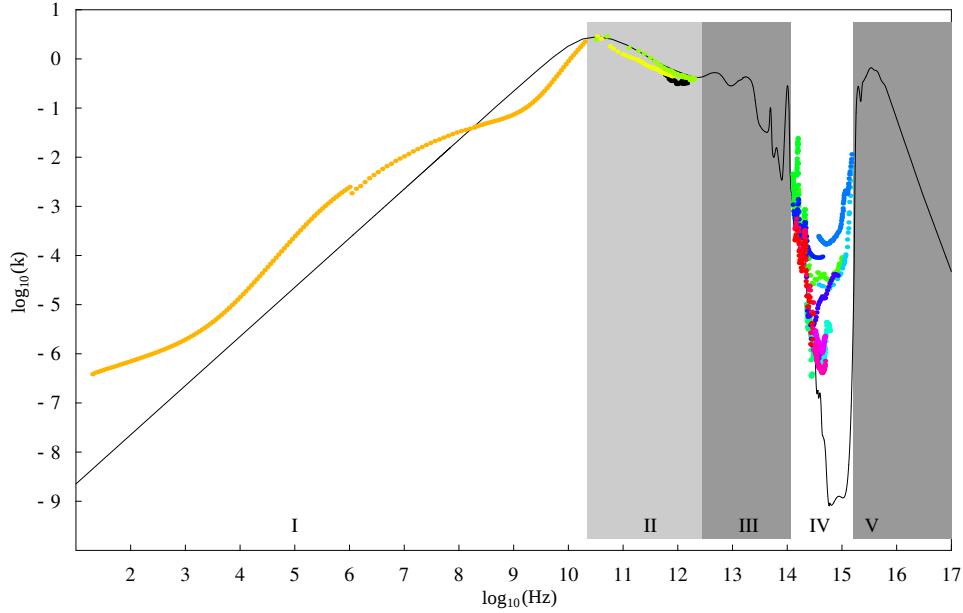}
\caption{Published absorption data for biological skin (colored points) and Segelstein's absorption curve for pure liquid water (solid black line). The width for each shaded region; Region I: DC to 22 GHz, Region II: 22 GHz to 3 THz, Region III: 3 THz to 2.5 $\mu$m (IR-B), Region IV: 2.5 $\mu$m (IR-B) to 189 nm (far-UV), Region V: 189 nm (far-UV) to x-ray.}\label{figure1}
\end{figure}

The Kramers-Kr\"{o}nig transform requires an absorption set $\mathcal{A}$ spanning the entire frequency spectrum be defined for biological skin. Surveying available experimental data for biological tissue reveals publication primarily in three frequency bands. Each experimental data point was digitized from published graphs, see colored points in Fig. (\ref{figure1}). Because of Corollary \ref{waterpostulate}, Segelstein's theoretical absorption curve for pure liquid water is included for immediate comparison \cite{Segelstein}.

Experimental conductivity data, ranging from 20 Hz to 20 GHz, supplied by one definitive source populates Region I, see orange colored points in Fig. (\ref{figure1}) \cite{Gabriel:AFRL}. It was found necessary to multiply calculated absorption ($k$) values by the vacuum permittivity ($\epsilon_0$) and angular frequency ($\omega$) after converting reported conductivity values ($\sigma'_{\textrm{exp}}$) using equation (\ref{complexrefraction}). The correction factor became evident after comparing calculated absorption values to that of pure liquid water. The additional correction factor is attributed to polarization of dielectric layers within skin samples. The corrected relation for experimental conductivity removes an additional dependency on capacitance from the Maxwell-Wagner effect ($\epsilon_0\omega\sigma'_{\textrm{exp}}=\epsilon''_r=\sigma'_r/\omega$) \cite{Pethig:dielectric,Martinsen,Barlea}. Gabriel's wet skin data confirms Corollary \ref{bound}, showing a contraction of the thermal resonance peak. An additional broad peak is seen centered about 6\logmine{Hz}, evidently from the influence of electrophoresis. Gabriel provides 171 data points with subinterval (1.2999, 10.3010)\logmine{Hz} to absorption set $\mathcal{A}$.
 
\begin{figure}[ht]
\centering
 \includegraphics[width=5.0 in]{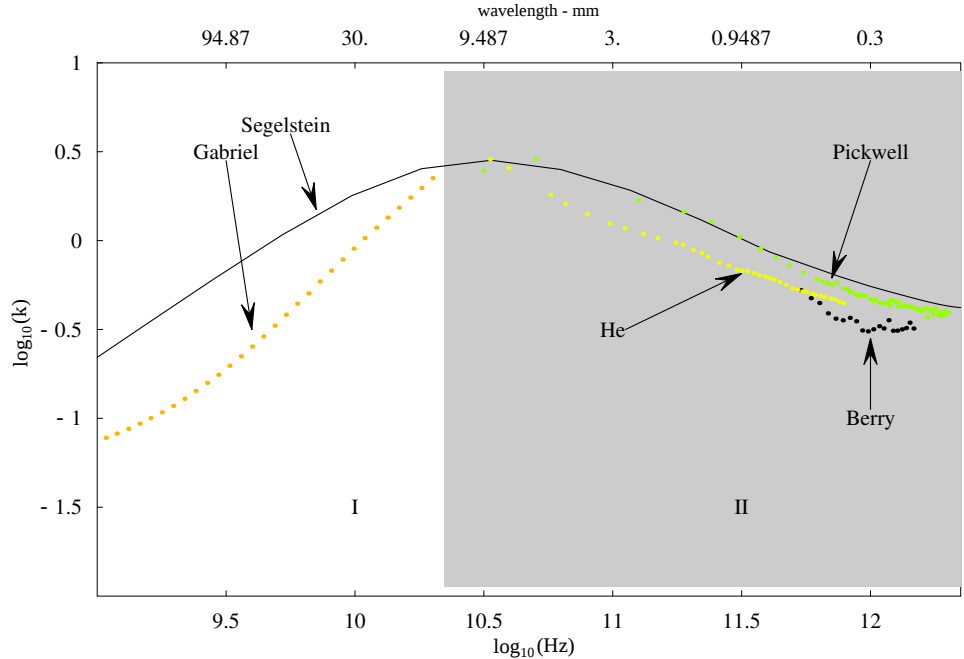}
\caption{Microwave-THz region: Compares reported absorption behavior of human skin (colored points) to pure liquid water (solid black line). Both Berry \etal\ and He \etal\ show a reduction in absorption for skin, while Pickwell \etal\ reported absorption values for human skin very close to pure liquid water.}\label{figure2}
\end{figure}

Because of the Similarity Theorem \ref{similarity} and reasons of symmetry, a similar reduction in absorption is expected in the THz band. Of the three authors reporting absorption for biological skin, 38 data points from He \etal\ reported for rat skin are added to absorption set $\mathcal{A}$, spanning subinterval (10.5235, 11.8942)\logmine{Hz} \cite{He}. Absorption values reported by Berry \etal\ for \emph{in vivo} human skin spanned too small an interval and did not characterize the most important area centered around the thermal maximum \cite{Berry}. Pickwell \etal\ employed a double Debye permittivity model to calculate reported absorption values for \emph{in vivo} human skin \cite{Pickwell}. The use of a double Debye permittivity model to extract optical constants returned values very similar to pure liquid water. Debye's permittivity model neglects intermolecular interactions and cannot represent contraction of the absorption peak \cite{McConnell}.

\begin{figure}[ht]
\centering
 \includegraphics[width=5.0 in]{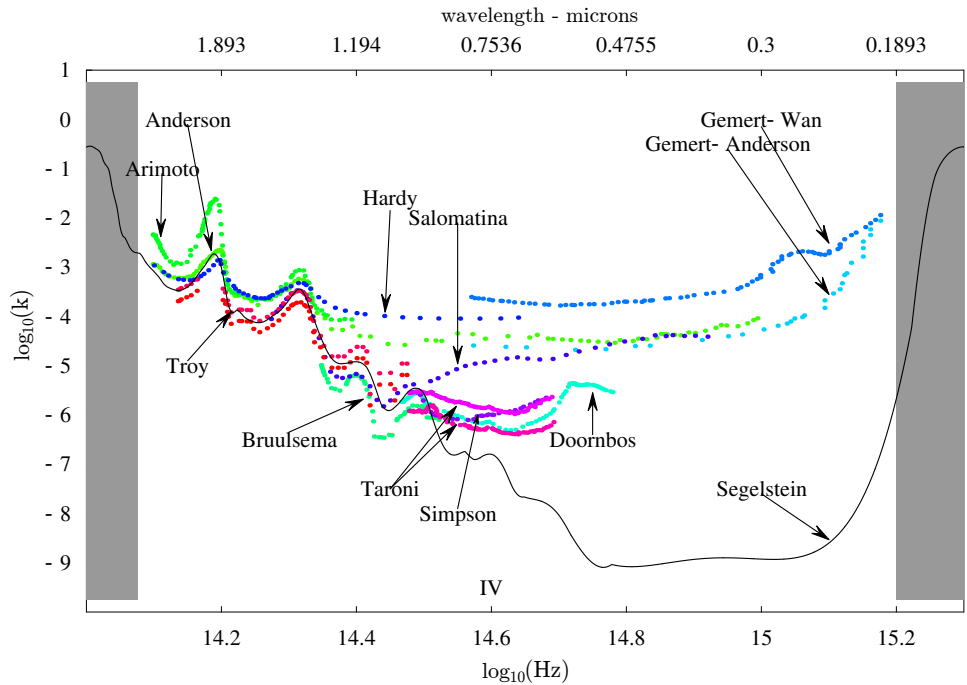}
\caption{IR-VIS-UV region: Compares absorption for human skin from eleven different authors (colored points) to Segelstein's water curve (solid black line). The use of SIS method and/or the Kubelka-Munk model leads to exceptionally high absorption.}\label{figure3}
\end{figure}

%Vogel \etal\ have stated the absorption behavior for biological tissue is expected to resemble water in the infrared spectra and diverge in the visible to ultraviolet spectra \cite{Vogel}. 
The presence of such scatterers as melanin, hemoglobin, and proteins are expected to raise absorption in the visible and ultraviolet spectra \cite{Vogel}. Yet, the absorption behavior for biological tissue is expected to resemble water in the infrared spectra. To a lesser or greater degree, this expectation is met by published experimental values collected, see Fig. (\ref{figure3}). Unfortunately, large discrepancies amongst reported values demand further discussion to discriminate which are valid. 

Pickering \etal\ found the single integrating sphere (SIS) method can suffer error 60\% higher than the double apparatus method \cite{Pickering}. The double integrating sphere method was shown to return non-unique results and fails for highly scattering media. Absorption values reported by Hardy \etal\ show the presence of two vibrational peaks, but the entire curve appears flat and non-responsive \cite{Hardy}. Hardy \etal\ used a derivative form of the single integrating sphere method, goniometry, where a reflective sphere is replaced by a detector on a rotating arm. 

In addition to error associated with experimental technique are numerical models used to calculate optical properties from reflectance and transmittance data. Yang \etal\ determined the widely used Kubelka-Munk (K-M) model fails for low absorptive media \cite{Yang}. Authors using both the SIS technique and K-M method report the highest absorption values, specifically, Gemert-Anderson and Gemert-Wan, where data was digitized from graphs published by Gemert \etal \cite{Gemert}. Salomatina \etal\ reports lower absorption for \emph{ex vivo} human skin, where optical constants were calculated using a Monte Carlo model in addition to using the SIS method \cite{Salomatina}. 

Troy \etal\ employed the more accurate double integrating sphere apparatus and reports absorption values for \emph{ex vivo} human skin in the infrared spectra  similar to pure liquid water \cite{Troy}. Arimoto \etal, Bruulsema \etal, Taroni \etal\ and Doornbos \etal\ used either time or spatially resolved reflectance spectroscopy methods on \emph{in vivo} human skin samples \cite{Arimoto,Bruulsema,Taroni,Doornbos}. Of particular interest are values reported by Simpson \etal, where a unique application of the SIS method appears to remedy the inherent inaccuracies \cite{Simpson}. Simpson \etal\ developed a unique calibrating scheme to correct for systematic errors and reports values similar to Troy \etal, Bruulsema \etal, Taroni \etal\ and Doornbos \etal.

In light of Postulate \ref{waterpostulate}, reported absorption for biological skin resembling pure liquid water cannot be accidental. All reported values agree skin's absorption diverges midway through the near-infrared spectra from pure liquid water, \emph{circa} 14.5\logmine{Hz}. After considering insights from both Pickering \etal\ and Yang \etal, reported absorption that is excessively high is judged invalid. The magnitude of skin's absorption in the visible spectra is on the order of -6\logmine{k} and will contribute very little to the Kramers-Kr\"{o}nig integral; therefore, wrangling over reported values is not warranted. 94 points of data reported by Doornbos \etal\ are added to set $\mathcal{A}$. This experimental set was chosen for two reasons, 1) the reported interval contains the point where skin's absorption diverges from pure liquid water and 2) the interval extends furthest into the visible spectra with range (14.4665, 14.7795)\logmine{Hz}. 

Because of similarities with pure liquid water and Postulate \ref{waterpostulate}, it is redundant to average reported absorption for biological skin in most of the infrared spectra. The vast expanse of the mid-infrared spectra is completely devoid of any experimental data, see shaded Region III in Fig. (\ref{figure1}). An intense literature search came up empty handed. In light of  reduced absorption in the THz band, it is not exactly clear where to place the left endpoint for supplanted water. In other words, approaching the THz band (He \etal) from the mid-infrared, skin's absorption must diverge at some point if it is to connect smoothly. 1741 data points from pure liquid water are added to set $\mathcal{A}$ with range (12.3198, 14.46063)\logmine{Hz}. The left endpoint of the supplanted water is arbitrary and interpolation is expected to supply needed data within the inserted void. 

The high frequency limit behavior, Corollary \ref{highlimit}, supports supplanting water absorption data for skin in the high frequency limit. To set $\mathcal{A}$ are added 20374 data points from water with range (15.2, 21)\logmine{Hz}. It is not known how to connect the right edge reported by Doorbos \etal\ with the left edge of the high frequency limit behavior. Both the single integrating sphere method and the Kubelka-Munk model were used to derive reported absorption in the ultraviolet spectra. Both methods have been determined inaccurate for low absorptive turbid media; consequently, another gap was left intentionally vacant in the experimental absorption set $\mathcal{A}$ with range [14.7795, 15.2]\logmine{Hz}. 

Theory states material absorption is null at the origin. The amplitude is vanishingly small for frequencies left of Gabriel's data ($\omega/2\pi<20$ Hz) and will not contribute greatly to the Kramers-Kr\"{o}nig integral. We join the origin to Gabriel's data using a simple linear regression model ($y=\gamma\omega$) that intercepts the origin, with calculated slope $\gamma=1.93041\times 10^{-8}$. 1,000 data points are appended to set $\mathcal{A}$ with frequency range (-32, 1.2999]\logmine{Hz}. The experimental absorption set $\mathcal{A}$ now possesses an overall range of (-32, 21)\logmine{Hz}. Interpolation is required to both generate a regular set of data and remove two large voids, specifically, the first void is near the THz band, range [11.8942, 12.3198]\logmine{Hz}, and the second within the ultraviolet spectra, range [14.7795, 15.2]\logmine{Hz}.
 
From set $\mathcal{A}$ a new set $\mathcal{A}'$ is created using a combination of interpolation followed by extrapolation. Approximations were made using Neville's interpolation algorithm for arbitrary frequency $\omega_0$. The interpolation order $p$ swept through all even integers 2n to produce a total of 30 approximations, where $n=\{1,2,\ldots,30\}$. Richardson's extrapolation algorithm, with order $2n=60$, was then successively applied to all 30 approximations to yield a single accurate approximation $\tilde{x}$, for frequency $\omega_0$. This process was repeated for all 10,000 chosen frequencies regularly spanning the experimental spectrum (-32, 21)\logmine{Hz}. Computational time was just under 9 hours to evaluate all 10,000 elements to form set $\mathcal{A}'$. The algorithm was applied in base ten logarithmic space to minimize numerical error.  
\begin{figure}[ht]
\centering
 \includegraphics[width=5.0 in]{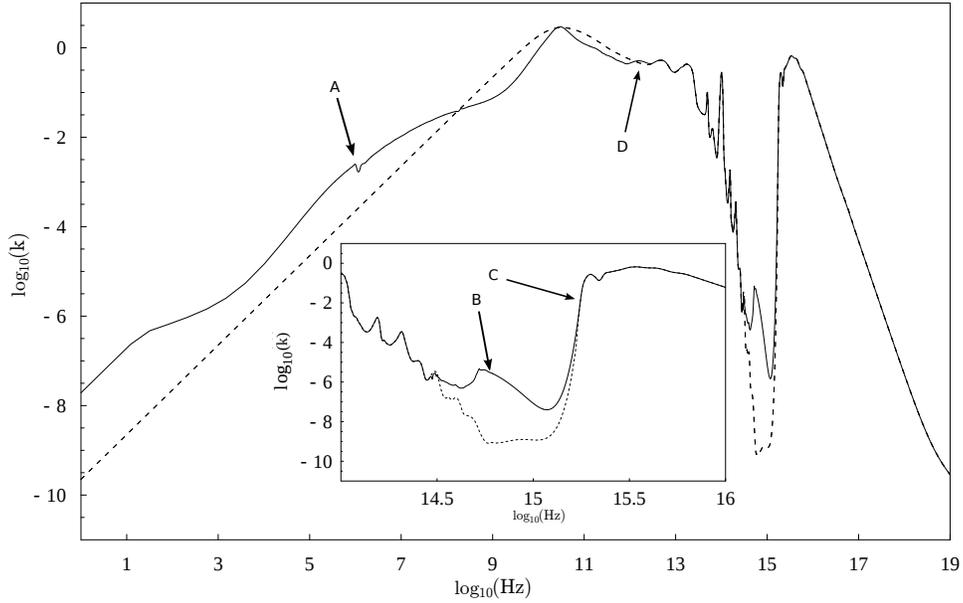}
\caption{Interpolated/extrapolated absorption set $\mathcal{A}'$ for biological skin (solid black line) and Segelstein's water absorption curve (dashed line). Inset focuses on the NIR-VIS-UV spectra. For details on lettered arrows see text below.}\label{figure4}
\end{figure}

Fig. (\ref{figure4}) shows the resulting absorption curve $\mathcal{A}'$ for biological skin after interpolation and extrapolation. An experimental artifact is retained from Gabriel's data, see arrow A in Fig. (\ref{figure4}). An extra peak is introduced by the interpolation scheme over the first void (arrow D). Interpolation filled the void intentionally left in the ultraviolet spectra, where the right endpoint of Doornbos \etal\ (arrow B) connects to the left endpoint of supplanted water values (arrow C). To test any possible ill effects in forming set $\mathcal{A}'$, measured relaxation times before and after interpolation/extrapolation are compared. Measuring the Full Width Half Maximum (FWHM) of the thermal peak from both absorption sets, $\mathcal{A}$ and $\mathcal{A}'$, returned the same approximate relaxation time, specifically 16.824 ps for skin. For comparison purposes, we measured the relaxation time from Segelstein's pure liquid water curve as 5.38 ps.

The Hilbert transform was used to emulate the Kramers-Kr\"{o}nig relation, see equation (\ref{KKeqn}). To satisfy Shannon's \emph{sampling theorem} for set $\mathcal{A}'$, a sampling rate of $5\times 10^{-22}$ is required which would create a set with $2\times10^{21}$ point density. This far exceeds available computing power and forced implementing two \emph{under-sampled} windows. The absorption magnitude is vanishingly small in the visible spectrum and provides a perfect place for the right endpoint of the first window, range (0, 14.8)\logmine{Hz}. It is satisfactory to place the edge of the second window far into the x-ray band where absorption is vanishingly small, range (0, 17)\logmine{Hz}. The windows are judiciously spaced to minimize boundary errors from the discrete Fourier transform (DFT).

Based on available computing power, a point density of $2^{21}$ was chosen for each window submitted. Proper implementation of the DFT requires a regularly sampled set of data points and each window shifted the placement and spacing of the resulting subintervals.  This requires another round of interpolations to be performed on set $\mathcal{A}'$. The spacing between each data point was considered sufficiently close to allow first order interpolation. Interpolations were performed in base ten logarithmic space utilizing Mathematica's built-in \textbf{Interpolation[]} algorithm \cite{Mathematica}. The first approximation is always taken one step-size from the origin. Interpolation produces a list $l_i$ with length $2^{21}$ for each window, where the subscript $i=\{1,2\}$ identifies which window is referenced. In order to represent absorption for negative frequencies, a new list $l'_i$ is created by taking the negative of list $l_i$ in reversed order. Combining lists $l_i$ and inverse lists $l'_i$ creates a composite list $\mathcal{L}_i$ representing skin absorption for both positive and  negative frequencies. Below is a concise definition of the composite list $\mathcal{L}_i$ using Mathematica's notation and functions:
\begin{equation}
 \mathcal{L}_i=\textrm{\textbf{Flatten[}}\{-\textrm{\textbf{Rest[Reverse[ }}l_i\textrm{\textbf{ ]]}},0,l_i\}\textrm{\textbf{]}}
\end{equation}

Theory states absorption is null at the origin, hence, the number zero is explicitly stated. The function \textbf{Flatten[]} removes all nested lists within the outer curly brackets. The function \textbf{Rest[]} removes one element from the list enclosed and this is to maintain an overall length for the composite list $\mathcal{L}_i$ equal to $2^{22}$. Lastly, the function \textbf{Reverse[]} obviously reverses the order of elements in the enclosed list. 

Using the definition for the sign function ($\sgn$), another list $\mathcal{L}_{\sgn}$ of length $2^{22}$ is created spanning both positive and negative frequencies, see equation (\ref{signfunction}). Using Mathematica's built-in \textbf{Fourier[]} function, the forward discrete Fourier transform ($\mathscr{F}$) is taken of the composite list $\mathcal{L}_i$ and then multiplied by list $\mathcal{L}_{\textrm{sign}}$. The discrete inverse Fourier transform ($\mathscr{F}^{-1}$) is taken of the product list, $\mathscr{F}^{-1}\hspace{-1pt}\{\textrm{\textbf{I}}\cdot\mathcal{L}_{\sgn}\times \mathscr{F}\hspace{-1pt}\{\mathcal{L}_i\}\}$, using Mathematica's built-in \textbf{InverseFourier[]} function, where \textbf{I} is the imaginary number $\sqrt{-1}$. Mathematica's discrete Fourier transform functions will implement Cooley and Tukey's \emph{Fast Fourier Transform} (FFT) algorithm for lists having length equal to an integer power of 2 \cite{Cooley}. The dispersion index is retrieved by adding unity to the list after the inverse Fourier transform is taken. Runtime was approximately 20 seconds to complete all numerical operations and process the 4,194,304 elements for each window. Half of the dispersion list generated is discarded, for we do not need the negative frequencies. 

Segelstein emphasized in his thesis that calculated dispersion curves are shape preserving, but not their amplitudes \cite{Segelstein}. This is a consequence of \emph{under-sampling}, which poorly represents the area under the absorption curve. The amplitude is commonly corrected by scaling calculated dispersion to an excepted static permittivity value for the material under study. If a series of windows are used, then the disperison index from the first window is scaled and each subsequent output is scaled to match the right endpoint of each preceding corrected output.

In the present study, the static permittivity is not known for biological skin. 
In the limit of static fields, the dispersion index is directly proportional to the square root of the real permittivity ($n^2=\epsilon'_r$), see equation (\ref{limitofnk}). It is common to see values with orders of magnitude 4, 5 and as high as 6 reported for skin's real permittivity  \cite{Gabriel:AFRL,Yamamoto,Schwan,Pethig:passive}. An estimate of the compound dielectric constant for heterogenous mixtures is attainable through the mixing rules and Maxwell-Garnett's formula \cite{Sihvola}. Pethig \etal\ reports static permittivity values of 90 and 110 for amino acids and proteins, respectively \cite{Pethig:passive}. Since water represents roughly 75 percent of tissue, fractional percentages for amino acids and proteins of 10 and 15 percent return a value of 83 for the effective permittivity. Gabriel reports a real permittivity ($\epsilon'_r$) of 56432 for wet skin at 20 Hz, this stands in stark contrast to estimates obtained through the mixing rules and Maxwell-Garnett's formula.

Because of the confusion concerning skin's static permittivity, another means of scaling dispersion index calculations is needed. It was noticed that predicted index values obtained from the second windowed transform compared favorably with index values reported by Ding \etal \cite{Ding}. Predicted values were within reported standard deviations for skin epidermis and dermis; moreover, the mean  value approached very close to calculated index values. In light of this discovery, it was decided to reverse the order of scaling by scaling the output from the first window to match an arbitrary point in the second windowed transform. 

\begin{figure}[ht]
\centering
 \includegraphics[width=5.0 in]{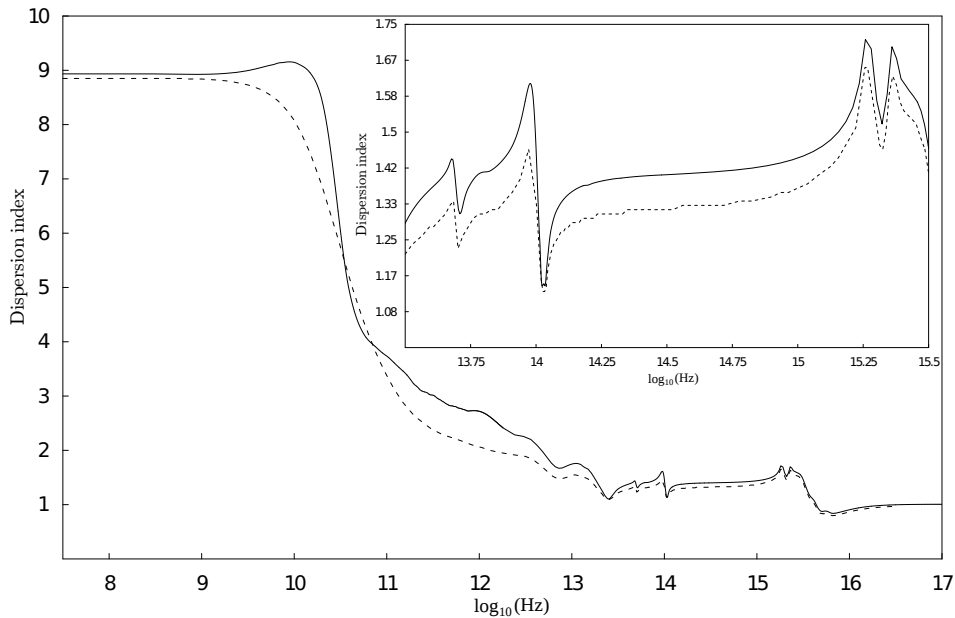}
\caption{Thoeretical dispersion index curve for both biological skin (solid line) and Segelstein's pure liquid water (dashed line). Both curves were generated through the Kramers-Kr\"{o}nig relation defining index as a function of absorption. Inset focuses upon IR-VIS-UV spectras.}\label{figure5}
\end{figure}

Wavelength 600 nm was arbitrarily chosen as the fixed point for scaling, the index predicted from the second windowed transform is approximately 1.41. This required multiplying the first windowed output by approximately 1.41 to raise the predicted value for 600 nm from unity. It is known that windowed transforms slope downward towards higher frequencies as a result of \emph{under-sampling} the Hilbert kernel, it was chosen to multiply rather than add the correction factor for this reason \cite{Bracewell}. The resulting dispersion curve is shown together with pure liquid water's theoretical dispersion curve for reference, see Fig. (\ref{figure5}).   

By inspection, the dispersion index for biological skin is not strictly greater than pure liquid water. The prominence of the dispersion curve around the thermal resonance verifies Corollary \ref{onetoone}, also, dilation of the relaxation time for bound water momentarily forces dispersion to dip slightly below that of pure liquid water, \emph{circa} 10.8\logmine{Hz}. The predicted static permittivity is $n^2\approx 80.5$ and compares well with estimates from Maxwell-Garnett's formula. The experimental absorption values collected are derived primarily from skin tissue samples, therefore the predicted dispersion is indicative of the combined index for the epidermis and dermis. Predicted index values compare very well with the mean index obtained from averaging dermis and epidermis index reported by Ding \etal \cite{Ding}.
%, except in one area, around 10\logmine{Hz}, where tissue is predicted to be less scattering than pure liquid water,
\section{Conclusion}
Kramers-Kr\"{o}nig analysis of the optical properties for biological skin produced a comprehensive set for the complex index of refraction spanning the practicable frequency spectrum, see Tables (\ref{table1}-\,\ref{table4}). Segelstein's theoretical curve for pure liquid water proved an invaluable resource in discriminating experimental absorption for biological skin. Through comparison, an additional correction factor was found necessary for experimental RF data; otherwise, the predicted slope was opposite to expected absorption behavior, increasing towards the origin instead of decreasing. Biological skin exhibits strong similarity to pure liquid water in the near-infrared spectra. Through comparison, derived absorption from the single integrating sphere method and Kubelka-Munk model were identified as exceptionally high and were not included in this analysis.  

It is surprising to discover results derived from widely accepted methods as the single integrating sphere method and Kubelka-Munk model were inaccurate. This should focus attention on the importance of global analysis, as well as, the need for further skin research in the mid-infrared and ultraviolet spectra. Of particular interest are the results reported by Simpson \etal, the unique experimental method he describes for the single integrating sphere method apparently makes viable what is otherwise an inaccurate method when applied to low absorptive turbid media.

A novel application of Richardson's extrapolation algorithm theoretically allowed greater accuracy for interpolated approximations. The artifact peak covering the intentional void near the THz band indicates a possible discontinuity in the tangent space between experimental skin data and supplanted water data. Due to a lack of experimental data, the exact behavior of biological skin is not known in the mid-infrared band. It is quite possible that we were too aggressive extending the left endpoint for supplanted water. In the case of the ultraviolet band, the interpolation/extrapolation algorithm produced a smooth regular curve connecting experimental data in the visible spectra with supplanted water used to represent the high frequency limit behavior. The implementation of the Kramers-Kr\"{o}nig transform requires only a simply connected curve. The absorption magnitude in the visible spectra is quite small and contributes little to the sum evaluated from the Kramers-Kr\"{o}nig integral. Even though the ultraviolet spectra is purely a product of interpolation, small changes in absorption values would not drastically change predictions found in this analysis. The area in the ultraviolet spectra is not as important as the area around the thermal maximum. The thermal resonance peak represents the area of largest contribution to the integral and it is fortunate to find experimental values for this region.   

Comparison of biological tissue with pure liquid water is based on the assumption water is a topological basis set $\mathcal{B}$. Postulate \ref{waterpostulate} imparts a global perspective and provides a more secure footing for analysis. With this concept, all possible biological tissue types are new topologies $\mathcal{X}$ created from the intersection of the basis set with other open sets ($s\in S$). Open sets containing elements such as proteins, amino acids, melanin and others give rise to new topologies. Ultimately, there are a family of curves based upon partial molar concentration, specific elements contained in open sets and the degree of intersection with the basis set. The manifold $\mathcal{M}$ of all possible tissue types is covered in a continuous fashion by the basis set and all included open sets. 

To fully characterize all possible tissue types or even biological skin alone would require a comprehensive analysis including the variation of all pertinent parameters. This would be a daunting task and it may be more efficacious to treat the manifold $\mathcal{M}$ as a probability space. Segelstein's theoretical curve brought all experimental data into proper focus. In like manner, it is hoped the published theoretical curve for biological skin will aid future investigations.

%%%%%%%%%%%%%%%%%%%%%%%%%%%%%%%%%%%%%%%%%%%%%%%%%%%%%%%%%%%%%
%\acknowledgments 
%This unnumbered section is used to identify those who have aided the authors in understanding or accomplishing the work presented and to acknowledge sources of funding.  

%%%%%%%%%%%%%%%%%%%%%%%%%%%%%%%%%%%%%%%%%%%%%%%%%%%%%%%%%%%%%
%%%%% References %%%%%

\bibliography{references}   %>>>> bibliography data in report.bib
\bibliographystyle{spiejour}   %>>>> makes bibtex use spiejour.bst

\pagebreak
%%first column
\begin{table}[ht]
\caption{The complex index of refraction for biological skin.}
\centering
\begin{tabular}{c c c | c c c}\label{table1}
$\textrm{Hz}^\textrm{\dag}$ & \textit{n} & $k^\textrm{\dag}$ & 
$\textrm{Hz}^\textrm{\dag}$ & \textit{n} & $k^\textrm{\dag}$ \\ \hline
 1.000000   &    8.939578   &   -6.714351   &    6.298990   &    8.935496   &   -2.482999   \\
 1.123232   &    8.939483   &   -6.591118   &    6.422222   &    8.935401   &   -2.374496   \\
 1.246465   &    8.939388   &   -6.467886   &    6.545455   &    8.935306   &   -2.281354   \\
 1.369697   &    8.939293   &   -6.377035   &    6.668687   &    8.935211   &   -2.186873   \\
 1.492929   &    8.939198   &   -6.331319   &    6.791919   &    8.935116   &   -2.109378   \\
 1.616162   &    8.939103   &   -6.287676   &    6.915152   &    8.935021   &   -2.029086   \\
 1.739394   &    8.939008   &   -6.244997   &    7.038384   &    8.934926   &   -1.957573   \\
 1.862626   &    8.938913   &   -6.199971   &    7.161616   &    8.934831   &   -1.878650   \\
 1.985859   &    8.938818   &   -6.155607   &    7.284848   &    8.934736   &   -1.810520   \\
 2.109091   &    8.938723   &   -6.109489   &    7.408081   &    8.934641   &   -1.744610   \\
 2.232323   &    8.938628   &   -6.062518   &    7.531313   &    8.934546   &   -1.681756   \\
 2.355556   &    8.938533   &   -6.014847   &    7.654545   &    8.934451   &   -1.630421   \\
 2.478788   &    8.938439   &   -5.964629   &    7.777778   &    8.934356   &   -1.572027   \\
 2.602020   &    8.938344   &   -5.911301   &    7.901010   &    8.934261   &   -1.519076   \\
 2.725253   &    8.938249   &   -5.853770   &    8.024242   &    8.934166   &   -1.472950   \\
 2.848485   &    8.938154   &   -5.791213   &    8.147475   &    8.934072   &   -1.434497   \\
 2.971717   &    8.938059   &   -5.722863   &    8.270707   &    8.933977   &   -1.405930   \\
 3.094949   &    8.937964   &   -5.648232   &    8.393939   &    8.933882   &   -1.347691   \\
 3.218182   &    8.937869   &   -5.566130   &    8.517172   &    8.933063   &   -1.303787   \\
 3.341414   &    8.937774   &   -5.475667   &    8.640404   &    8.930668   &   -1.265113   \\
 3.464646   &    8.937679   &   -5.376275   &    8.763636   &    8.928273   &   -1.227062   \\
 3.587879   &    8.937584   &   -5.268221   &    8.886869   &    8.926229   &   -1.179096   \\
 3.711111   &    8.937489   &   -5.150880   &    9.010101   &    8.926533   &   -1.122481   \\
 3.834343   &    8.937394   &   -5.024149   &    9.133333   &    8.931372   &   -1.050549   \\
 3.957576   &    8.937299   &   -4.890220   &    9.256566   &    8.943282   &   -0.961375   \\
 4.080808   &    8.937204   &   -4.749078   &    9.379798   &    8.962965   &   -0.847667   \\
 4.204040   &    8.937109   &   -4.602029   &    9.503030   &    8.996557   &   -0.716077   \\
 4.327273   &    8.937014   &   -4.450784   &    9.626263   &    9.044252   &   -0.563216   \\
 4.450505   &    8.936920   &   -4.295949   &    9.749495   &    9.096353   &   -0.392385   \\
 4.573737   &    8.936825   &   -4.140328   &    9.872727   &    9.142406   &   -0.216169   \\
 4.696970   &    8.936730   &   -3.983998   &    9.995960   &    9.147571   &   -0.043311   \\
 4.820202   &    8.936635   &   -3.828108   &    10.119190   &    9.043294   &    0.119454   \\
 4.943434   &    8.936540   &   -3.674416   &    10.242420   &    8.752549   &    0.277131   \\
 5.066667   &    8.936445   &   -3.524499   &    10.365660   &    7.854737   &    0.419840   \\
 5.189899   &    8.936350   &   -3.378856   &    10.488890   &    6.196389   &    0.468220   \\
 5.313131   &    8.936255   &   -3.238912   &    10.612120   &    4.809049   &    0.392951   \\
 5.436364   &    8.936160   &   -3.106452   &    10.735350   &    4.191943   &    0.278733   \\
 5.559596   &    8.936065   &   -2.982193   &    10.858590   &    3.938295   &    0.174631   \\
 5.682828   &    8.935970   &   -2.866537   &    10.981820   &    3.758090   &    0.097715   \\
 5.806061   &    8.935875   &   -2.759004   &    11.105050   &    3.571038   &    0.042165   \\
 5.929293   &    8.935780   &   -2.658342   &    11.228280   &    3.367261   &   -0.005768   \\
 6.052525   &    8.935685   &   -2.765799   &    11.351520   &    3.129285   &   -0.076384   \\
 6.175758   &    8.935590   &   -2.587471   &    11.474750   &    3.021477   &   -0.158801   \\ \hline
\end{tabular}
\\ $^\textrm{\dag}$ base ten logarithm\\
\end{table}

\pagebreak
%%second column
\begin{table}[ht]
\caption{The complex index of refraction for biological skin.}
\centering 

\begin{tabular}{ccc|ccc}\label{table2}
$\textrm{Hz}^\textrm{\dag}$ & \textit{n} & $k^\textrm{\dag}$ & 
$\textrm{Hz}^\textrm{\dag}$ & \textit{n} & $k^\textrm{\dag}$ \\\hline
 11.597980   &    2.906948   &   -0.206702   &    13.523620   &    1.314257   &   -1.399415   \\
 11.721210   &    2.811804   &   -0.279284   &    13.535180   &    1.324821   &   -1.418134   \\
 11.844440   &    2.745357   &   -0.325889   &    13.546730   &    1.334406   &   -1.434744   \\
 11.967680   &    2.729055   &   -0.353438   &    13.558290   &    1.343177   &   -1.449080   \\
 12.090910   &    2.646382   &   -0.308833   &    13.569850   &    1.351247   &   -1.460799   \\
 12.214140   &    2.475084   &   -0.291899   &    13.581410   &    1.358595   &   -1.470832   \\
 12.337370   &    2.319472   &   -0.325557   &    13.592960   &    1.365624   &   -1.480685   \\
 12.460610   &    2.258284   &   -0.357444   &    13.604520   &    1.372462   &   -1.487215   \\
 12.583840   &    2.163877   &   -0.299143   &    13.616080   &    1.379076   &   -1.491992   \\
 12.707070   &    1.927734   &   -0.281131   &    13.627640   &    1.386022   &   -1.494465   \\
 12.830300   &    1.687664   &   -0.387174   &    13.639200   &    1.393793   &   -1.490207   \\
 12.953540   &    1.718446   &   -0.545354   &    13.650750   &    1.403163   &   -1.471619   \\
 13.076770   &    1.751993   &   -0.478728   &    13.662310   &    1.415433   &   -1.425704   \\
 13.200000   &    1.607991   &   -0.382958   &    13.673870   &    1.434151   &   -1.295379   \\
 13.211560   &    1.580444   &   -0.373971   &    13.685430   &    1.425557   &   -1.048490   \\
 13.223120   &    1.549564   &   -0.369169   &    13.696980   &    1.353495   &   -1.010719   \\
 13.234670   &    1.519690   &   -0.366611   &    13.708540   &    1.310461   &   -1.240625   \\
 13.246230   &    1.487561   &   -0.366588   &    13.720100   &    1.328862   &   -1.647880   \\
 13.257790   &    1.453021   &   -0.367824   &    13.731660   &    1.356099   &   -1.904144   \\
 13.269350   &    1.420910   &   -0.373040   &    13.743220   &    1.372916   &   -1.979483   \\
 13.280900   &    1.388183   &   -0.378204   &    13.754770   &    1.384852   &   -2.003717   \\
 13.292460   &    1.354386   &   -0.387560   &    13.766330   &    1.394403   &   -1.971577   \\
 13.304020   &    1.322328   &   -0.398473   &    13.777890   &    1.401472   &   -1.905777   \\
 13.315580   &    1.289478   &   -0.410752   &    13.789450   &    1.405677   &   -1.841590   \\
 13.323230   &    1.267251   &   -0.421281   &    13.801010   &    1.407429   &   -1.806282   \\
 13.327140   &    1.256401   &   -0.427063   &    13.812560   &    1.407913   &   -1.824808   \\
 13.338690   &    1.221609   &   -0.446150   &    13.824120   &    1.409823   &   -1.901921   \\
 13.350250   &    1.186705   &   -0.474133   &    13.835680   &    1.414206   &   -2.002140   \\
 13.361810   &    1.155860   &   -0.510963   &    13.847240   &    1.419732   &   -2.104408   \\
 13.373370   &    1.132535   &   -0.556592   &    13.858790   &    1.426159   &   -2.209069   \\
 13.384920   &    1.108097   &   -0.615093   &    13.870350   &    1.433532   &   -2.310891   \\
 13.396480   &    1.106628   &   -0.698015   &    13.881910   &    1.441882   &   -2.396236   \\
 13.408040   &    1.109368   &   -0.786359   &    13.893470   &    1.451527   &   -2.453692   \\
 13.419600   &    1.129723   &   -0.890640   &    13.905030   &    1.463097   &   -2.458168   \\
 13.431160   &    1.155407   &   -0.983932   &    13.916580   &    1.477733   &   -2.356314   \\
 13.442710   &    1.178894   &   -1.071893   &    13.928140   &    1.496482   &   -2.141512   \\
 13.446460   &    1.186614   &   -1.102451   &    13.939700   &    1.520595   &   -1.849001   \\
 13.454270   &    1.203519   &   -1.166657   &    13.951260   &    1.550530   &   -1.550473   \\
 13.465830   &    1.230831   &   -1.244639   &    13.962810   &    1.584446   &   -1.251301   \\
 13.477390   &    1.253400   &   -1.295537   &    13.974370   &    1.610859   &   -0.972857   \\
 13.488940   &    1.273150   &   -1.331567   &    13.985930   &    1.596208   &   -0.735036   \\
 13.500500   &    1.288986   &   -1.356378   &    13.997490   &    1.496570   &   -0.580602   \\
 13.512060   &    1.302464   &   -1.378905   &    14.009050   &    1.316883   &   -0.555232   \\ \hline
\end{tabular}
\\ $^\textrm{\dag}$ base ten logarithm\\
\end{table}

\pagebreak
%%Third column
\begin{table}[ht]
\caption{The complex index of refraction for biological skin.}\vspace{2pt}
\centering
\begin{tabular}{ccc|ccc}\label{table3}
$\textrm{Hz}^\textrm{\dag}$ & \textit{n} & $k^\textrm{\dag}$ & 
$\textrm{Hz}^\textrm{\dag}$ & \textit{n} & $k^\textrm{\dag}$ \\\hline
 14.020600   &    1.151473   &   -0.691072   &    14.517590   &    1.401684   &   -5.877865   \\
 14.032160   &    1.144321   &   -1.032114   &    14.529150   &    1.402076   &   -5.931121   \\
 14.043720   &    1.205133   &   -1.705126   &    14.540700   &    1.402468   &   -5.979235   \\
 14.055280   &    1.267127   &   -2.365138   &    14.552260   &    1.402862   &   -6.029363   \\
 14.066830   &    1.298337   &   -2.649755   &    14.563820   &    1.403259   &   -6.103398   \\
 14.078390   &    1.317689   &   -2.725881   &    14.575380   &    1.403660   &   -6.144097   \\
 14.089950   &    1.330899   &   -2.933979   &    14.586930   &    1.404068   &   -6.163130   \\
 14.101510   &    1.341168   &   -3.109568   &    14.598490   &    1.404482   &   -6.154727   \\
 14.113070   &    1.349229   &   -3.309102   &    14.610050   &    1.404905   &   -6.252584   \\
 14.124620   &    1.355872   &   -3.430778   &    14.621610   &    1.405339   &   -6.302773   \\
 14.136180   &    1.361342   &   -3.470704   &    14.633170   &    1.405784   &   -6.304187   \\
 14.147740   &    1.365984   &   -3.417124   &    14.644720   &    1.406242   &   -6.245585   \\
 14.159300   &    1.369994   &   -3.265796   &    14.656280   &    1.406715   &   -6.173099   \\
 14.170850   &    1.373354   &   -3.051120   &    14.667840   &    1.407204   &   -6.113629   \\
 14.182410   &    1.375924   &   -2.819953   &    14.679400   &    1.407712   &   -6.032218   \\
 14.193970   &    1.376597   &   -2.783803   &    14.690950   &    1.408238   &   -5.907989   \\
 14.205530   &    1.378498   &   -3.409379   &    14.702510   &    1.408787   &   -5.661629   \\
 14.217090   &    1.381229   &   -3.905255   &    14.714070   &    1.409358   &   -5.413335   \\
 14.228640   &    1.383242   &   -3.932203   &    14.725630   &    1.409952   &   -5.384350   \\
 14.240200   &    1.385005   &   -4.067073   &    14.737190   &    1.410575   &   -5.386847   \\
 14.251760   &    1.386585   &   -4.122733   &    14.748740   &    1.411228   &   -5.388691   \\
 14.263320   &    1.388013   &   -4.102375   &    14.760300   &    1.411912   &   -5.442384   \\
 14.274870   &    1.389304   &   -4.000567   &    14.771860   &    1.412632   &   -5.509158   \\
 14.286430   &    1.390479   &   -3.884667   &    14.783420   &    1.413388   &   -5.550228   \\
 14.297990   &    1.391551   &   -3.704063   &    14.794970   &    1.414185   &   -5.601480   \\
 14.309550   &    1.392428   &   -3.500757   &    14.806530   &    1.415024   &   -5.658709   \\
 14.321110   &    1.392995   &   -3.510090   &    14.818090   &    1.415910   &   -5.720890   \\
 14.332660   &    1.393758   &   -3.896329   &    14.829650   &    1.416846   &   -5.787299   \\
 14.344220   &    1.394593   &   -4.316987   &    14.841210   &    1.417835   &   -5.857507   \\
 14.355780   &    1.395358   &   -4.679023   &    14.852760   &    1.418882   &   -5.931209   \\
 14.367340   &    1.396052   &   -4.902601   &    14.864320   &    1.419990   &   -6.008159   \\
 14.378890   &    1.396684   &   -4.957659   &    14.875880   &    1.421166   &   -6.088263   \\
 14.390450   &    1.397264   &   -4.936130   &    14.887440   &    1.422413   &   -6.171450   \\
 14.402010   &    1.397798   &   -4.924163   &    14.898990   &    1.423738   &   -6.257634   \\
 14.413570   &    1.398292   &   -5.016344   &    14.910550   &    1.425147   &   -6.346766   \\
 14.425130   &    1.398756   &   -5.337472   &    14.922110   &    1.426646   &   -6.438502   \\
 14.436680   &    1.399193   &   -5.765644   &    14.933670   &    1.428243   &   -6.532454   \\
 14.448240   &    1.399600   &   -5.909267   &    14.945230   &    1.429947   &   -6.628085   \\
 14.459800   &    1.399979   &   -5.806298   &    14.956780   &    1.431766   &   -6.724667   \\
 14.471360   &    1.400332   &   -5.760596   &    14.968340   &    1.433710   &   -6.821246   \\
 14.482910   &    1.400663   &   -5.587046   &    14.979900   &    1.435790   &   -6.916569   \\
 14.494470   &    1.400970   &   -5.662902   &    14.991460   &    1.438020   &   -7.009190   \\
 14.506030   &    1.401291   &   -5.768644   &    15.003020   &    1.440412   &   -7.097395   \\ \hline
\end{tabular}
\\ $^\textrm{\dag}$ base ten logarithm\\
\end{table}

\pagebreak
%%Fourth column
\begin{table}[ht]
\caption{The complex index of refraction for biological skin.}\vspace{2pt}
\centering
\begin{tabular}{ccc|ccc}\label{table4}
$\textrm{Hz}^\textrm{\dag}$ & \textit{n} & $k^\textrm{\dag}$ & 
$\textrm{Hz}^\textrm{\dag}$ & \textit{n} & $k^\textrm{\dag}$ \\\hline
 15.014570   &    1.442984   &   -7.179190   &    15.545450   &    1.255985   &   -0.183690   \\
 15.026130   &    1.445753   &   -7.252298   &    15.590910   &    1.127595   &   -0.227846   \\
 15.037690   &    1.448739   &   -7.313698   &    15.636360   &    1.016363   &   -0.240617   \\
 15.049250   &    1.451967   &   -7.360688   &    15.681820   &    0.892072   &   -0.332410   \\
 15.060800   &    1.455463   &   -7.390154   &    15.727270   &    0.880039   &   -0.454941   \\
 15.072360   &    1.459259   &   -7.398627   &    15.772730   &    0.867331   &   -0.516173   \\
 15.083920   &    1.463393   &   -7.382357   &    15.818180   &    0.839561   &   -0.630444   \\
 15.095480   &    1.467909   &   -7.336948   &    15.863640   &    0.847511   &   -0.777584   \\
 15.107040   &    1.472861   &   -7.257415   &    15.909090   &    0.866336   &   -0.924420   \\
 15.118590   &    1.478315   &   -7.140005   &    15.954550   &    0.886338   &   -1.069408   \\
 15.130150   &    1.484350   &   -6.979974   &    16.000000   &    0.905396   &   -1.218095   \\
 15.141710   &    1.491071   &   -6.772384   &    16.045450   &    0.922860   &   -1.366243   \\
 15.153270   &    1.498608   &   -6.512135   &    16.090910   &    0.937885   &   -1.513756   \\
 15.164820   &    1.507138   &   -6.192159   &    16.136360   &    0.950634   &   -1.660817   \\
 15.176380   &    1.516903   &   -5.808037   &    16.181820   &    0.961241   &   -1.805593   \\
 15.187940   &    1.528256   &   -5.354897   &    16.227270   &    0.969916   &   -1.949286   \\
 15.199500   &    1.541735   &   -4.827317   &    16.272730   &    0.977038   &   -2.096410   \\
 15.211060   &    1.558258   &   -4.219894   &    16.318180   &    0.982914   &   -2.245046   \\
 15.222610   &    1.579596   &   -3.526534   &    16.363640   &    0.987729   &   -2.391733   \\
 15.234170   &    1.609840   &   -2.739866   &    16.409090   &    0.991627   &   -2.537240   \\
 15.245730   &    1.658328   &   -1.870447   &    16.454550   &    0.994790   &   -2.684529   \\
 15.257290   &    1.711574   &   -1.173216   &    16.500000   &    0.997371   &   -2.828810   \\
 15.268840   &    1.721371   &   -0.834522   &    16.545450   &    0.999446   &   -2.966014   \\
 15.280400   &    1.694463   &   -0.658089   &    16.590910   &    1.001109   &   -3.099469   \\
 15.291960   &    1.636985   &   -0.573362   &    16.636360   &    1.002442   &   -3.231468   \\
 15.303520   &    1.563346   &   -0.568237   &    16.681820   &    1.003512   &   -3.363734   \\
 15.315080   &    1.524880   &   -0.639451   &    16.727270   &    1.004366   &   -3.497451   \\
 15.326630   &    1.520918   &   -0.740412   &    16.772730   &    1.005045   &   -3.633330   \\
 15.338190   &    1.566765   &   -0.854803   &    16.818180   &    1.005576   &   -3.771598   \\
 15.349750   &    1.641636   &   -0.830620   &    16.863640   &    1.005980   &   -3.912050   \\
 15.361310   &    1.701895   &   -0.663869   &    16.909090   &    1.006268   &   -4.053048   \\
 15.372860   &    1.687771   &   -0.519489   &    16.954550   &    1.006438   &   -4.188535   \\
 15.384420   &    1.643915   &   -0.463465   &    17.000000   &    1.006096   &   -4.324606   \\
 15.395980   &    1.622511   &   -0.448992   &      &      &      \\
 15.407540   &    1.612289   &   -0.426236   &      &      &      \\
 15.419100   &    1.602458   &   -0.401595   &      &      &      \\
 15.430650   &    1.591922   &   -0.377467   &      &      &      \\
 15.442210   &    1.583239   &   -0.352683   &      &      &      \\
 15.453770   &    1.573387   &   -0.324545   &      &      &      \\
 15.465330   &    1.559044   &   -0.293864   &      &      &      \\
 15.476880   &    1.537025   &   -0.262668   &      &      &      \\
 15.488440   &    1.505400   &   -0.233148   &      &      &      \\
 15.500000   &    1.462397   &   -0.208048   &      &      &      \\ \hline 
\end{tabular}
\\ $^\textrm{\dag}$ base ten logarithm\\
\end{table}

\end{document}